\documentclass[%
 pra,
 amsmath,amssymb,
reprint,%
superscriptaddress,
floatfix,
]{revtex4-2}

\usepackage{graphicx,float}
\usepackage{dcolumn}
\usepackage{bm,braket}
\usepackage{dsfont}
\usepackage[utf8]{inputenc}
\usepackage[T1]{fontenc}
\usepackage{mathptmx}
\usepackage{tikz}
\usetikzlibrary{positioning}
\usepackage{siunitx,booktabs}
\usepackage{hyperref}
\usepackage{soul}
\begin{document}


\title[Draft]{Improved Efficiency of Open Quantum System Simulations Using Matrix Products States in the Interaction Picture}

\author{Hanggai Nuomin}%

\affiliation{ 
Department of Chemistry, Duke University, Durham, North Carolina 27708, United States
}%
\author{David N. Beratan}%
\email{david.beratan@duke.edu.}
\affiliation{ 
	Department of Chemistry, Duke University,
	Durham, North Carolina 27708, United States
}\affiliation{ 
	Department of
	Biochemistry, Duke University,
	Durham, North Carolina 27710, United States
}\affiliation{ 
	Department of Physics, Duke University,
	Durham, North Carolina 27708, United States
}
\author{Peng Zhang}%
\email{peng.zhang@duke.edu.}
\affiliation{ 
	Department of Chemistry, Duke University, Durham, North Carolina 27708, United States
}%


\date{\today}

\begin{abstract}
Modeling open quantum systems---quantum systems coupled to a bath---is of value in condensed matter theory, cavity quantum electrodynamics, nanosciences and  biophysics. The real-time simulation of open quantum systems was advanced significantly by the recent development of chain mapping techniques and the use of matrix product states that exploit the intrinsic entanglement structure in open quantum systems. The computational cost of simulating open quantum systems, however, remains high when the bath is excited to high-lying quantum states.
We develop an approach to reduce the computational costs in such cases. The interaction representation for the open quantum system is used to distribute excitations among the bath degrees of freedom so that the occupation of each bath oscillator is ensured to be low. The interaction picture also causes the matrix dimensions to be much smaller in a matrix product state of a chain-mapped open quantum system than in the Schr\"odinger picture. Using the interaction-representation accelerates the calculations by 1-2 orders of magnitude over existing matrix-product-state method. In the regime of strong system-bath coupling and high temperatures, the speedup can be as large as 3 orders of magnitude. The approach developed here is especially promising to simulate the dynamics of open quantum systems in the high-temperature and strong-coupling regimes \cite{ivander2021strong}.

\end{abstract}

\maketitle

\section{Introduction}
\label{sec:intro}
The density matrix renormalization group (DMRG) method \cite{white1992density} has become a standard tool to study low-dimensional systems, especially one and two-dimensional spin systems \cite{white1993density,yan2011spin}. The essence of the DMRG method is to retain the most significant states associated with the relevant properties of a quantum system and discard the less important ones using singular values decomposition. The time-dependent version of DMRG (t-DMRG) \cite{white2004real,daley2004time,xie2019time} is promising to describe quantum dynamics, and it can  achieve high accuracy at moderate computational cost. The DMRG and t-DMRG methods are most easily understood using the language of the matrix-product-state (MPS) \cite{vidal2003efficient,affleck1987rigorous,affleck1987rigorousCMP,schollwock2011density} representation of quantum states.

\begin{figure}
	\includegraphics[width=\linewidth]{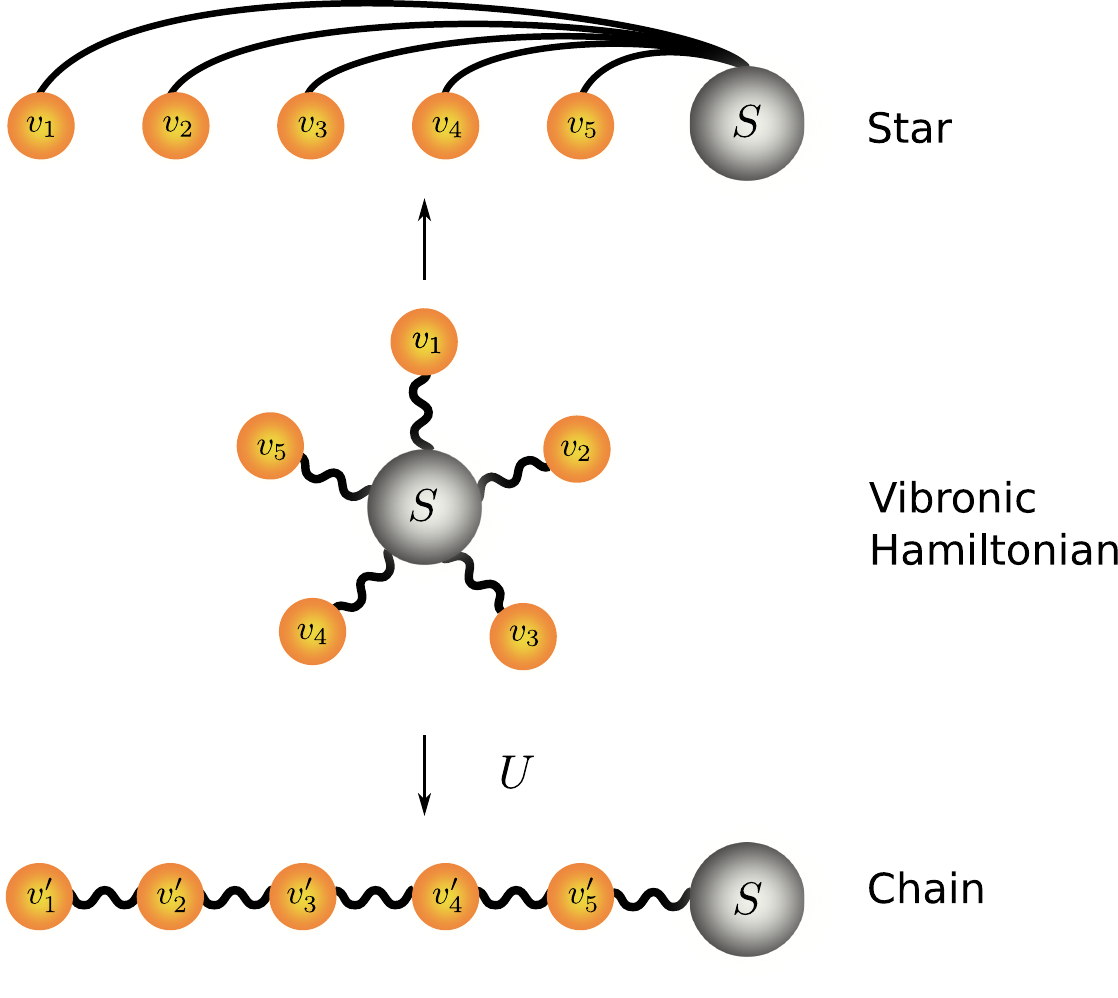}
	\caption{Configurations of the star and chain geometries. The lines represent the interactions between sites. The star geometry does not change the vibronic Hamiltonian \emph{per se} and it is a 1D representation of the Hamiltonian. The ordering of bath modes is crucial to the efficiency of the star geometry. Ordering by the magnitudes of bath-mode frequencies is commonly used \cite{borrelli2021finite}. The chain Hamiltonian is obtained from the vibronic Hamiltonian by a unitary transformation $U$ and it contains interaction terms between two adjacent bath modes.}
	\label{fig:chainAndStar}
\end{figure}

\begin{figure}
	\includegraphics[width=\linewidth]{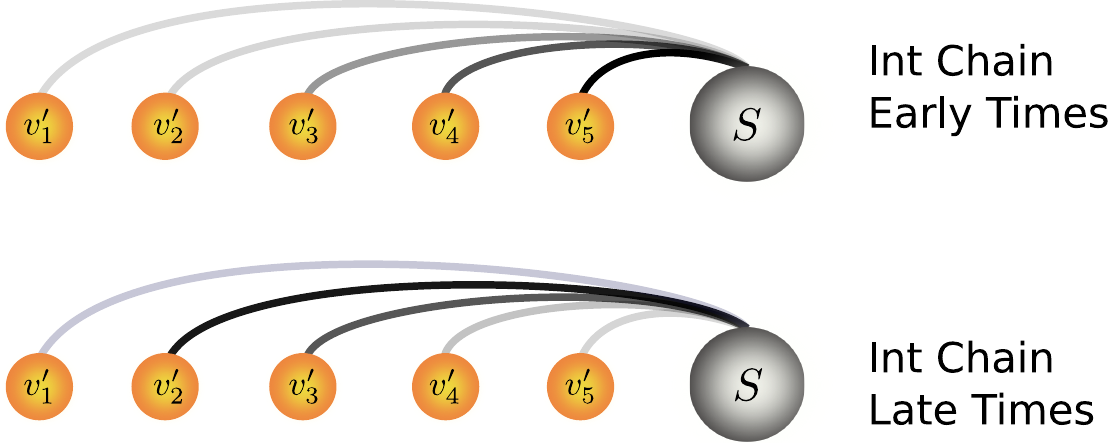}
	\caption{Configuration of the interaction-picture chain Hamiltonian. In the interaction-picture chain Hamiltonian, the mode-mode terms are removed and the system interacts with all of the modes. An important feature of the interaction-picture Hamiltonian is that the interaction strengths of the system and bath modes are time-dependent and spatially  localized. Here, we use the thickness of lines to indicate the interaction strengths. At early times, the system interacts strongly with the first several modes. At later times, the strongest interactions move to  further parts of the 1D chain. The interactions are always localized in  a narrow band of vibrational modes (see Fig.~\ref{fig:dk}). In contrast, in the star-geometry, the interactions are not spatially localized.}
	\label{fig:geometries}
\end{figure}

The MPS representation was developed to describe one-dimensional systems, including the Ising model and its  variants \cite{white1993density} and the Bose-Hubbard model \cite{kuhner2000one}. Using t-DMRG to study the condensed-phase quantum dynamics of electron or energy transfer, for example, requires that the electron-vibration Hamiltonian be mapped onto one dimension, although the electron-vibration Hamiltonian has an intrinsic star-shaped topology \cite{huo2011theoretical,landry2012recover}---the electronic system interacts with all of the vibrational modes present (see Fig.~\ref{fig:chainAndStar}). The requirement of t-DMRG for a 1D configuration is fulfilled by arranging the bath modes along a line and placing the electronic system at the end of the line, as shown in Fig.~\ref{fig:chainAndStar}. This configuration is known as the star geometry \cite{stoudenmire2010minimally,wall2016simulating}. Using the star geometry does not change the content of an electron-vibration Hamiltonian,  and it places the electronic system and the bath modes into a one-dimension topology, introducing long-range interactions between the electronic system and the bath modes (Fig.~\ref{fig:geometries}). Despite the long-range interactions, the star geometry was proven to be efficient when  calculating  the Green's functions of a special open quantum system (for example, the Anderson impurity models where the impurity, a magnetic atom, is the quantum system and the conduction electrons are the bath) at zero temperature in the context of MPS \cite{wolf2014solving}.

For open quantum systems other than the impurity model, the star geometry is not necessarily efficient numerically, because it involves long-range interactions.  
Long-range interactions are believed to make MPS simulations computationally expensive because the bond dimensions (i.e., the number of important singular values of the density matrix) in a MPS grows, in general, rapidly as a function of time. In addition, the effect of finite temperatures on the numerical efficiency of the star geometry and the chain geometry is not clear. The recently-developed  approach of using a thermalized spectral density \cite{tamascelli2019efficient,de2015thermofield} to describe finite temperature open quantum systems maps a finite-temperature bath to an effective zero-temperature bath, and it is not clear whether the star geometry remains efficient in this thermo-field description of finite-temperature effects. A strategy that is  different from using the star geometry, the chain geometry mapping  \cite{prior2010efficient,de2015discretize}, was proposed to avoid the long-range interactions present in the star geometry. The chain geometry differs from the star geometry in the topology of the  bath modes (vibrations). In the chain geometry, the bath modes are mapped to a 1D chain of oscillators with nearest-neighbor interactions, and the quantum system interacts only with the first bath mode in the 1D chain. The short-time dynamics of open quantum systems can be described largely by a few modes in close proximity to the system \cite{sanchez2021accurate}.
Many quantum dynamical simulations have been performed using the chain geometry, and it is believed that the chain geometry usually (has a lower computational cost) than the star geometry since the bond dimensions of MPS in the chain geometry are small (because the long-range interactions are absent).

Although the bond dimensions in the chain geometry are believed to be smaller than in the star geometry, the computational cost of open quantum system simulations with a boson bath also depends on another quantity---the number of eigenstates of each bath mode (i.e., the \emph{local dimension}). When $k_BT$ ($T$ is temperature) is large compared to the energy spacing of the bath modes, the local dimension for each bath mode must be large enough  to ensure  numerical convergence of the simulations. Similarly, in the strong system-bath coupling regime, large local dimensions for the bath modes are also necessary for numerical convergence. The large local dimensions may slow simulations that use a chain geometry. For example, in the chain geometry, if the dimension of the Hilbert space for a bath mode (local dimension) is 100 (see Sec.~\ref{sec:res} for an example), an interaction term $\kappa_n(\hat{b}_n^\dagger\hat{b}_{n-1}+\mathrm{h.c.})$, where $b^\dagger_{n}$ is the creation operator for the $n$-th mode in the chain,  generates an evolution {operator} with a {corresponding} matrix of the size $10^4\times10^4$, causing time-consuming matrix operations. Furthermore, large local dimensions of bath modes also significantly affect the size of the tensors in a MPS because the dimension of {one of the indices} in a MPS tensor is equal to the local dimension of the corresponding site (here, a bath mode). The large tensors in MPS then make the required singular value decomposition a difficult task. In summary, when using the chain geometry, the dimensions of the evolution matrices or {of} the Hamiltonian matrices become large at high temperatures or in the strong coupling regime, and the corresponding numerical simulations in these regimes become impractical.
This large-matrix problem does not exist in the star geometry  because the system is often described by a smaller Hilbert space than the bath modes, and the interaction terms in the star geometry couple the system to each bath mode, making the matrix size much smaller than in the chain geometry. However, the bond dimensions in the star geometry grow rapidly with time since long-range interactions induce strong entanglement, which makes long-time simulations challenging. The small dimension of the interaction terms in the star geometry motivate us to transform the interaction terms in the chain geometry to resolve the large-matrix difficulty that impedes the chain-geometry simulations,  especially in the regimes of high temperatures and strong system-bath interactions.

Now, we describe an approach to overcome the computational challenges of the chain and star geometries by transforming the chain Hamiltonian to the interaction picture. In the interaction picture, the transformed Hamiltonian does not contain the mode-mode interaction terms, thus producing much smaller evolution matrices. The transformed chain Hamiltonian features a star geometry with time-dependent system-bath couplings. In contrast to the star geometry in the Schr\"odinger picture, the time-dependent couplings are spatially localized, producing slower growth of entanglement.

We use the transformed chain geometry in the interaction picture to describe the spin-boson model and compare the resulting dynamics, bond dimensions, and computational costs to the treatment with the Schr\"odinger-picture chain and star geometries.

The paper is organized as follows. Section~\ref{sec:theory} derives the full and the truncated (finite chain length) form of the chain-geometry Hamiltonian in the interaction picture. The truncated form is used for the spin-boson model simulations. Section~\ref{sec:res} compares the results obtained with the chain-geometry Hamiltonian in the interaction picture, the chain geometry in the the Schr\"odinger picture and the star geometry
.
These studies demonstrate the speed and accuracy of the interaction-picture chain geometry approach. Using scaling analysis, we show that the interaction-picture chain geometry approach is 1-2 orders of magnitude faster than the other two schemes if the vibrational mode excitations are high. The strategy developed here allows the use of t-DMRG and tensor networks to simulate strongly coupled vibronic systems in the high-temperature regime ($k_BT\gg \hbar\omega_c$ where $\omega_c$ is the characteristic frequency of the bath), such as the strong-coupling-enabled non-thermal coupled electron-exciton transport (quantum ratcheted reactions \cite{bhattacharyya2020quantum}), and ambient-temperature excitation energy transfer reactions, for example. This method is also suitable to model ultra-strong coupling and deep-strong coupling regimes (for a description of these regimes, see Ref \cite{wang2015bridging,moroz2014hidden}) in light-matter interactions \cite{forn2019ultrastrong}. Section~\ref{sec:conclusion} summarizes features of the computational scheme (chain geometry in the interaction picture) and discusses possible improvements to this scheme.

\section{THEORY AND METHODOLOGY}
\label{sec:theory}
\subsection{Orthogonal Polynomial Mapping of the Hamiltonian}
The  Hamiltonian describes a quantum system coupled to a bath (i.e., an open quantum system) with $\hbar=1$,
\begin{eqnarray}
	H =&&H_{s}+ H_{sb} + H_{b}\\
	  =&& H_{s} + A_{s}\otimes \int_{\Omega_0}^{\Omega_1}(\hat{a}_\omega^\dagger + \hat{a}_\omega)h(\omega)\,\mathrm{d}\omega \nonumber \\
	   &&+ \int_{\Omega_0}^{\Omega_1}\hat{a}_\omega^\dagger \hat{a}_\omega \omega \,\mathrm{d}\omega
	  \label{eq:star}
\end{eqnarray}
$\hat{a}^\dagger_\omega$ and $\hat{a}_\omega$ are the creation and annihilation operators for a boson mode of frequency $\omega$. $h(\omega)$ is the coupling strength between the system and the bath mode of frequency $\omega$. The frequencies $\Omega_0$ and $\Omega_1$ are the upper and lower limits of the bath frequencies. When $T>\SI{0}{\kelvin}$,  the spectral density has a temperature-dependent form $J(\omega, \beta)=\mathrm{sign}(\omega)J(|\omega|)\left[1+\coth\left(\beta\omega/2\right)\right]/2$ with $\beta=1/k_BT$, and the lower limit $\Omega_0$ is extended to $-\infty$ because negative-frequency bath modes are needed to construct the Boltzmann distribution of bath states \cite{tamascelli2019efficient,dunnett2020simulating}.  Thus, at finite temperatures, $h(\omega)$ also becomes temperature-dependent:  $h(\omega,\beta)=\sqrt{J(\omega,\beta)}$. When the temperature is \SI{0}{\kelvin}, $h(\omega)=\sqrt{J(\omega)}$ {where $J(\omega)$ is the spectral density} with $0<\omega<\infty$ because no negative-frequency bath modes are needed to construct the Boltzmann distribution.

The Hamiltonian {in} Eq.~\eqref{eq:star} with a continuous distribution of bath frequencies can be transformed to {a} discrete nearest-neighbor form (Eq.~\eqref{eq:mapped}) by using a set of real, continuous, orthogonal and normalized polynomials $p_n(\omega)$ generated by the weight function $h^2(x)$ \cite{chin2010exact,prior2010efficient} to construct a unitary transformation $U$ with elements $U_n(\omega)=h(\omega)p_n(\omega)$.
\begin{eqnarray}
	\tilde{H} =&&H_{s} + \underbrace{\kappa_0A_{s}\otimes(\hat{b}_0+\hat{b}_0^\dagger)}_{\tilde{H}_{sb}} \nonumber \\
	&&+
	\underbrace{\sum_{n=0}^{\infty}{\omega}_nb_n^\dagger \hat{b}_n+\sum_{n=1}^{\infty}\kappa_n (\hat{b}_n^\dagger \hat{b}_{n-1}+\mathrm{h.c.})}_{\tilde{H}_{b}}. \label{eq:mapped}
\end{eqnarray}
where $\hat{b}_n$ is a linear combination of $\hat{a}_n$. In the transformed Hamiltonian, the system only interacts with one transformed vibrational mode, the zero-th mode ($\hat{b}_0,\hat{b}_0^\dagger$). 

The linear transformation $U$ relates $\hat{a}_\omega^\dagger$ and $\hat{b}_{n}^\dagger$:
\begin{equation} \label{eq:transformU}
\hat{a}_\omega^\dagger = \sum_{n=0}^{\infty}U_{n}(\omega) \hat{b}_{n}^\dagger=\sum_{n=0}^{\infty}h(\omega)p_n(\omega) b_n^\dagger
\end{equation}
where $p_n(\omega)$ is the set of polynomials and $\hat{b}^\dagger_n$ and $\hat{b}_n$ are the creation and annihilation operators for mode $n$ of the chain. The orthonormal polynomials $p_n(\omega)$ satisfy the recurrence relation
\begin{eqnarray}
	p_{n+1}(x) = (C_n x-A_n)p_{n}(x)-B_np_{n-1}(x), \quad n=0,1,2\ldots \label{eq:recurrence}
\end{eqnarray}
with $p_{-1}(x)=0$ \cite{chin2010exact,rosenbach2016numerical}. The  coefficients $A_n,B_n,C_n$ are related to the coupling strength $\kappa_n$ and frequency $\omega_n$ \cite{chin2010exact}: 
\begin{eqnarray}
	A_n = \frac{\omega_n}{\kappa_{n+1}}, B_n=\frac{\kappa_{n}}{\kappa_{n+1}}, C_n = \frac{1}{\kappa_{n+1}},\quad n=0,1,2 \ldots.
\end{eqnarray}

Using the recurrence relation (Eq.~\eqref{eq:recurrence}), we can {compute} $p_n(x)$ given the frequencies $\omega_n$, couplings $\kappa_n$ and the first two polynomials $p_{-1}(x)=0,  p_{0}=1/\kappa_0=1/\sqrt{\int_{\Omega_0}^{\Omega_1}h^2(x)\, \mathrm{d}x}$. The  recurrence relations, the {relationships among} $\omega_n,\kappa_n, A_n,B_n$ and $C_n$, and the properties of the orthogonal polynomials, are described in Refs \cite{prior2010efficient,chin2010exact}.

\subsection{Chain Hamiltonian in the Interaction Picture}
 $\tilde{H}$ in Eq.~\eqref{eq:mapped} is the chain Hamiltonian. $H$ in Eq.~\eqref{eq:star} is the star Hamiltonian.
 The time evolution of the system and bath is governed by $\tilde{H}$ or $H$. In the Sch\"odinger picture, the time evolution is
\begin{eqnarray}
	\ket{\psi(t)}_S=e^{-i\tilde{H}t}\ket{\psi(0)}_S.
\end{eqnarray}
$\tilde{H}$ contains couplings between two adjacent bath modes. By transforming the chain Hamiltonian to the interaction picture, the coupling terms between adjacent bath modes are removed, and only the system-bath {coupling}s and the system operator remain, as we will show. The dimensions of the Hamiltonian matrices that describe these system-bath coupling terms are much smaller than the mode-mode coupling terms in the Schr\"odinger picture, since the dimension of the system Hilbert space is often small \cite{bian2021modeling,wu2021electronic,segal2010numerically} (equal to 2 for a spin system, for example, while the dimension of a single bath mode can be around 100 \cite{brockt2015matrix}). Therefore, we expect that the computational cost of simulating the interaction-picture Hamiltonian in Eq.~\eqref{eq:mapped} with MPS can be much lower than that in the Schr\"odinger-picture chain geometry. 

\subsubsection{The Chain Hamiltonian in the Interaction Picture} 
In the interaction picture with respect to $\tilde{H}_{b}$, the wave function is $\ket{\psi(t)}_I=e^{i\tilde{H}_{b}t}\ket{\psi(t)}_S$, and the evolution operator is $\hat{U}_I(t,0)=e^{i\tilde{H}_{b}t}e^{-i\tilde{H}t}$ which has the time evolution
\begin{eqnarray}
	\frac{\mathrm{d}}{\mathrm{d}t}\hat{U}_I(t,0) 
	&&= e^{i\tilde{H}_{b}t}(-i\tilde{H}+i\tilde{H}_{b})e^{-i\tilde{H}t}\nonumber\\
	&&=e^{i\tilde{H}_{b}t}(-i\tilde{H}+i\tilde{H}_{b})e^{-i\tilde{H}_{b}t}e^{i\tilde{H}_{b}t}e^{-i\tilde{H}t}\\
	&&=-ie^{i\tilde{H}_{b}t}({H}_{s}+\tilde{H}_{sb})e^{-i\tilde{H}_{b}t}\hat{U}_I(t,0).
\end{eqnarray}
The interaction-picture Hamiltonian $\tilde{H}_I$ is $e^{i\tilde{H}_{b}t}({H}_{s}+\tilde{H}_{sb})e^{-i\tilde{H}_{b}t}$.

It is difficult to begin with $e^{i\tilde{H}_{b}t}({H}_{s}+\tilde{H}_{sb})e^{-i\tilde{H}_{b}t}$ to obtain an expansion of $\tilde{H}_I$ in terms of $\hat{b}_n$ and $\hat{b}_n^\dagger$. Instead, we first turn to the star geometry and define the interaction-picture star-geometry Hamiltonian $H_I(t)$ (Eq.~\eqref{eq:star}) as
\begin{eqnarray}
	H_I(t) &&= e^{i{H}_{b}t}({H}_{s}+{H}_{sb})e^{-i{H}_{b}t}\nonumber\\
	    &&= H_{s} + A_{s}\otimes \int_{\Omega_0}^{\Omega_1}(\hat{a}_\omega^\dagger e^{i\omega t} + \hat{a}_\omega e^{-i\omega t})h(\omega)\,\mathrm{d}\omega.
	    \label{eq:star-int}
\end{eqnarray}
$\tilde{H}_I(t)$ and $H_I(t)$ are related by the chain mapping transformation $U_n$ (Eq.~\eqref{eq:transformU}) \cite{prior2010efficient}. Therefore,
\begin{widetext}
	\begin{eqnarray}
		\tilde{H}_I(t)
		&&= H_{s} + A_{s}\otimes \int_{\Omega_0}^{\Omega_1}\left[\sum_{n=0}^{\infty}U_{n}(\omega) \hat{b}_{n}^\dagger e^{i\omega t} + \sum_{n=0}^{\infty}U_{n}(\omega) \hat{b}_{n} e^{-i\omega t}\right]h(\omega)\,\mathrm{d}\omega\nonumber.
		\\
		&&= H_{s} + A_{s}\otimes\sum_{n=0}^{\infty}\left[ \hat{b}_{n}^\dagger\int_{\Omega_0}^{\Omega_1}h^2(\omega)p_{n}(\omega)  e^{i\omega t}\,\mathrm{d}\omega + \hat{b}_{n}\int_{\Omega_0}^{\Omega_1}h^2(\omega)p_{n}(\omega)  e^{-i\omega t}\,\mathrm{d}\omega
		\right]
		\\
		&&= H_{s} + A_{s}\otimes\sum_{n=0}^{\infty}\left[ d^*_n(t)\hat{b}_{n}^\dagger + d_n(t)\hat{b}_{n}
		\right],\label{eq:chain-int}
	\end{eqnarray}
Here, $d_n(t)=\int_{\Omega_0}^{\Omega_1}h^2(\omega)p_{n}(\omega)  e^{-i\omega t}\,\mathrm{d}\omega$ and $d^*_n(t)$ is the complex conjugate of $d_n(t)$. By definition, $d_n(t)$ is the Fourier transform of $p_n(\omega)$ with the weight function $h^{2}(\omega)$ for $\Omega_0<\omega<\Omega_1$.
\end{widetext}

Using $\tilde{H}_I(t)$, the evolution operator $\hat{U}(t,0)$ in the interaction picture is $\hat{U}(t,0)=e^{-i\int_{0}^{t} \tilde{H}(s)\,\mathrm{d}s}$ and $\hat{U}(t+\delta t,t)=e^{-i\int_{t}^{t+\delta t} \tilde{H}_I(s)\,\mathrm{d}s}$. The time integral in the exponential is $a(\omega,t,\delta t)=\int_{t}^{t+\delta t}e^{-i\omega s}\,\mathrm{d}s 
$. The evolution operator on the time interval $[t, t+ \delta t]$ {is}
\begin{equation}
	U(t+\delta t, t)=\exp\{-i[H_s \delta t+A_{s}\otimes \sum_{n=0}^{\infty}(d_n(\omega,t,\delta t )\hat{b}_{n} +\mathrm{h.c.})]\}
\end{equation}
where $d_n(\omega,t,\delta t )=\int_{\Omega_0}^{\Omega_1}h^2(\omega)p_{n}(\omega) a(\omega,t,\delta t)\,\mathrm{d}\omega$ and $\delta t$ is the size of a single evolution step. 

\subsubsection{Truncated Form of the Chain Hamiltonian in the Interaction Picture}
The Hamiltonian in Eq.~\eqref{eq:chain-int} describes a semi-infinite chain. In simulations, the semi-infinite chain will be truncated in one of two ways. One approach  is to keep the first $N$ terms in the summation of Eq.~\eqref{eq:chain-int}. 

The other truncation scheme is easier in the sense that the integrals in Eq.~\ref{eq:chain-int} are not needed. It begins with the truncated form of the original \emph{chain} Hamiltonian (this Hamiltonian is denoted as the C form) Eq.~\eqref{eq:mapped}
\begin{eqnarray}
	\tilde{H} =&&H_{s} + \underbrace{\kappa_0A_{s}\otimes(\hat{b}_0+\hat{b}_0^\dagger)}_{\tilde{H}_{sb}} \nonumber \\
	&&+
	\underbrace{\sum_{n=0}^{N}{\omega}_nb_n^\dagger \hat{b}_n+\sum_{n=1}^{N}\kappa_n (\hat{b}_n^\dagger \hat{b}_{n-1}+\mathrm{h.c.})}_{\tilde{H}_b}. \label{eq:mapped-trunc}
\end{eqnarray} 
To transform this Hamiltonian into the interaction picture with respect to $\tilde{H}_b$, we first diagonalize $\tilde{H}_b$. The diagonalization uses the fact that $\tilde{H}_b$ can be written
\begin{eqnarray}
	\tilde{H}_b = \mathbf{b^\dagger} \mathbf{M} \mathbf{b} 
\end{eqnarray}
where $\mathbf{\hat{b}^\dagger}=(\hat{b}_0^\dagger, \ldots, \hat{b}_N^\dagger)$, $\mathbf{b}=(\hat{b}_0, \ldots, \hat{b}_N)^T$ and the Lanczos tridiagonal matrix \cite{de2015discretize} $\mathbf{M}$ is
\begin{eqnarray}
	\mathbf{M} = 
	\begin{pmatrix}
		\omega_0 & \kappa_1 &                &  & \\
		\kappa_1 & \omega_1 & \kappa_2       &        & \\
		         & \ddots   & \ddots         & \ddots  &\\
                 &          & \kappa_{N-1}   & \omega_{N-1}  &\kappa_{N}\\
		         &          &                & \kappa_N  &\omega_{N}\\
	\end{pmatrix}.
\end{eqnarray}
$\mathbf{M}$ can be {diagonalized}  by the unitary transformation: $\mathbf{M} = \mathbf{P}^\dagger \bm{\Lambda} \mathbf{P}$ where $\bm{\Lambda}=\mathrm{diag}(\lambda_0,\ldots,\lambda_N)$ is a diagonal matrix with the elements {equal to} the frequencies of the independent {bath} modes. Defining $\mathbf{a}=(\hat{a}_0,\ldots,\hat{a}_k,\ldots,\hat{a}_N)^T=\mathbf{P}\cdot\mathbf{b}$ which implies $\mathbf{b}=\mathbf{P}^\dagger \cdot \mathbf{a}$, the Hamiltonian $\tilde{H}$ becomes $H$, that is, the \emph{star} Hamiltonian (denoted as the S form)
\begin{eqnarray}
	H=H_{s} + A_{s}\otimes\sum_{k=0}^{N}\kappa_0 (P_{0,k}^\dagger\hat{a}_k+P_{k,0}\hat{a}_k^\dagger)+	\sum_{k=0}^{N}{\lambda}_ka_k^\dagger \hat{a}_k
	\label{eq:star-obtained-from-chain}
\end{eqnarray}
where $P_{k,0}$ is the first column of $\mathbf{P}$ and $P_{0,k}^\dagger$ is the first row of $\mathbf{P}^\dagger.$

The next steps are to transform the Hamiltonian (Eq.~\eqref{eq:star-obtained-from-chain}) to the interaction picture {with respect to} the bath Hamiltonian $\sum_{k=0}^{N}{\lambda}_ka_k^\dagger \hat{a}_k$ and then {write} the ladder operators $\hat{a}_k, \hat{a}_k^\dagger$ {in terms of} the chain geometry {ladder operators}, namely $\hat{b}_n$ and $\hat{b}_n^\dagger$. The resulting truncated star Hamiltonian in the interaction picture is 
\begin{eqnarray} \label{eq:HIstar}
 	{H}_I =&&H_{s} + A_{s}\otimes\sum_{k=0}^{N}\kappa_0 (P_{0,k}^\dagger e^{-i\lambda_k t} \hat{a}_k + P_{k,0}e^{i\lambda_k t} \hat{a}_k^\dagger )
\end{eqnarray}
Using $\hat{a}_k=\sum_{n=0}^{N}P_{k,n} \hat{b}_n $ and $ \hat{a}^\dagger_k=\sum_{n=0}^{N}P_{n,k}^\dagger \hat{b}_n^\dagger $ to transform $\hat{a}_k, \hat{a}_k^\dagger$ to $\hat{b}_n$ and $\hat{b}_n^\dagger$ in Eq.~\eqref{eq:HIstar} leads to the chain-geometry Hamiltonian in the interaction picture (denoted as the IC form),
\begin{eqnarray} 
	\tilde{H}_I =&&H_{s} + A_{s}\otimes\sum_{k=0}^{N}\sum_{n=0}^{N}\kappa_0 (P_{0,k}^\dagger P_{k,n} e^{-i\lambda_k t} \hat{b}_n + \mathrm{h.c.} )\\
	             =&&H_{s} + A_{s}\otimes\sum_{n=0}^{N}(d_n(t) \hat{b}_n + \mathrm{h.c.} )\\
	             =&&H_s + H_{s,0}(t) + H_{s,1}(t)  + \cdots + H_{s,N}(t)
	\label{eq:IC}
\end{eqnarray}
where we define $d_n(t)=\sum_{k=0}^{N}\kappa_0 P_{0,k}^\dagger P_{k,n} e^{-i\lambda_k t}$. In Eq.~\eqref{eq:IC}, we also defined $H_{s,n}(t)=A_{s}\otimes [d_n(t)\hat{b}_n+\mathrm{h.c.}]$. This truncated Hamiltonian of Eq.~\eqref{eq:IC} in the interaction picture is used in the simulations of Section~\ref{sec:res}.

The one time step evolution operator for the IC Hamiltonian (Eq.~\eqref{eq:IC}) is 
\begin{eqnarray}
	\hat{U}(t+\delta t, t)=&&e^{-i[H_s\delta t + \sum_{n=0}H_{s,n}(t+\delta t/2)\delta t ]}\\
	                      =&&e^{-i[H_s \delta t+A_{s}\otimes \sum_{n=0}^{N}(d_n(t+\delta t/2) \hat{b}_{n }\delta t + \mathrm{h.c.})]}.
    \label{eq:single-evo}
\end{eqnarray}
Here, we use the midpoint Hamiltonian $H_{s,n}(t+\delta t/2)$
to approximate the time-dependent Hamiltonians on the interval $[t,t+\delta t]$.

The evolution operator in Eq.~\eqref{eq:single-evo} is then Trotterized, for example, to  second order in the time step,
\begin{eqnarray}
	U(t+\delta t, t) \approx &&e^{-i[H_s+A_{s}\otimes (d_0(t+\delta t/4)\hat{b}_{0} +\mathrm{h.c.})] \delta t/2}\nonumber\\
	                 &&e^{-i[A_{s}\otimes (d_1(t+\delta t/4)\hat{b}_{1} +\mathrm{h.c.})] \delta t/2}\cdots\nonumber\\
	                 &&e^{-i[A_{s}\otimes (d_N(t+\delta t/2)\hat{b}_{N} +\mathrm{h.c.})] \delta t/2}\nonumber\\
	                 &&e^{-i[A_{s}\otimes (d_N(t+\delta t/2)\hat{b}_{N} +\mathrm{h.c.})] \delta t/2}\cdots\nonumber\\
	                 &&e^{-i[A_{s}\otimes (d_1(t+\delta t/4)\hat{b}_{1} +\mathrm{h.c.})] \delta t/2}\nonumber\\
	                 &&e^{-i[H_s+A_{s}\otimes (d_0(t+\delta t/4)\hat{b}_{0} +\mathrm{h.c.})] \delta t/2}.
	                 \label{eq:chain-int-tro}
\end{eqnarray}
The advantage of {using} the interaction picture evolution operator (Eq.~\eqref{eq:chain-int-tro}) is that the dimensions of the evolution matrices are much smaller than those in the original  Schr\"odinger picture for the chain geometry. The original Schr\"odinger-picture chain Hamiltonian Eq.~\eqref{eq:mapped-trunc} contains terms, such as $\hat{b}_n\hat{b}_{n-1}$, and the dimension of $\hat{b}_n$ and $\hat{b}_{n-1}$ has to be large enough to ensure convergence. In the interaction picture, in contrast, the Hamiltonian only contains $H_{s}$ and interaction terms of the form $A_{sys}\otimes (d_n(t)\hat{b}_n+\mathrm{h.c.})$. {Since} the dimension of the system is often much smaller than the dimension of the bath vibrational modes ($\hat{b}_n,\hat{b}_n^\dagger$), the sizes of the evolution operator matrices in the interaction picture are much smaller than in the Schr\"odinger picture for the chain.

\subsection{Time Evolution of Matrix Product States}
After obtaining the chain Hamiltonian and the corresponding evolution operator in the interaction picture, one is ready to use a matrix product state to represent a wave function and apply the evolution operator to the MPS to compute dynamics. We briefly review the use of matrix products states and one of the time evolution algorithms for MPS employed here, namely the time-evolution block-decimation method  (TEBD)  \cite{vidal2003efficient,shi2006classical}.

Given a state $\ket{\psi(\sigma_s,\sigma_0,\ldots,\sigma_N)}$ {with} $N+2$ spatially {localized} states $\ket{\sigma_s}$,  $\ket{\sigma_0}$, $\ket{\sigma_1}$, $\ldots$, $\ket{\sigma_N}$ {on} sites $s,0,1,\ldots,N$, respectively (here $s$ {denotes} a system), the MPS representation of this state is
\begin{eqnarray}
	\ket{\psi} = \sum_{\{\sigma_i\}}\sum_{\{a_j\}} \Gamma_{a_0}^{\sigma_s}S_{a_0}\Gamma_{a_0,a_1}^{\sigma_0}\cdots S_{a_{N}}\Gamma_{a_{N}}^{\sigma_N}\ket{\sigma_1\cdots\sigma_{N}}.
	\label{eq:MPS}
\end{eqnarray}
Here $\{\sigma_i\}$ contains $\sigma_s,\sigma_0,\ldots,\sigma_N$ and $\{a_j\}$ contains $a_0,\ldots,a_{N}$. We refer to the labels of the sites' local states $\sigma_s,\sigma_0,\ldots,\sigma_{N}$ as physical legs. The diagonal matrices ${S_{a_i}}$ in Eq.~\eqref{eq:MPS} are the singular values obtained by Schmidt decomposing the state $\ket{\psi}$ into a sum of direct product states in two sub-spaces, namely the two sub-spaces spanned by $\ket{\sigma_s,\ldots,\sigma_i}$ and $\ket{\sigma_{i+1},\ldots,\sigma_{N}}$, respectively.

The MPS state can be most easily evolved if the Hamiltonian of the system has a nearest-neighbor form, using a Trotterized evolution operator. However, in the Hamiltonian Eq.~\eqref{eq:IC}, the interaction pattern does not {satisfy} the nearest-neighbor form because the Hamiltonian includes long-range interactions. To evolve the state using the star Hamiltonian, one can use swap gates \cite{stoudenmire2010minimally,biamonte2017tensor} to exchange positions of adjacent sites. A swap gate exchanges the two physical legs of two adjacent tensors. This exchange can be combined with an individual time evolution term and does not introduce additional computational cost if one orders the individual evolution terms in the Trotter expansion of the entire evolution operator properly. The evolution operator in Eq.~\eqref{eq:IC} {is appropriately} ordered. To show how the swap gates and evolution operators are combined, we write Eq.~\eqref{eq:chain-int-tro-swap} in the form used in practical calculations. That is, the evolution operator, Eq.~\eqref{eq:IC}, has swap gates inserted between all pairs of adjacent individual terms, except for the two $H_{s,N}$ terms. 
\begin{eqnarray}
	 &&U(t+\delta t, t) \nonumber=\\
	 &&\hat{S}_{0,s}e^{-i[H_s+H_{s,0}(t+\delta t/4)]\delta t/2}\nonumber\\
	 &&\hat{S}_{1,s}e^{-iH_{s,1}(t+\delta t/4)\delta t/2}\hat{S}_{2,s}e^{-iH_{s,2}(t+\delta t/4)\delta t/2} \nonumber\\
	 &&\hat{S}_{3,s}e^{-iH_{s,3}(t+\delta t/4)\delta t/2}\cdots\hat{S}_{N-1,s}e^{-iH_{s,N-1}(t+\delta t/4)\delta t/2}\nonumber \\
	 &&e^{-iH_{s,N}(t+\delta t/4)\delta t/2}e^{-iH_{s,N}(t+\delta t/4)\delta t/2}\nonumber\\
	 &&S_{s,N-1}e^{-iH_{s,N}(t+\delta t/4)\delta t/2}\cdots \hat{S}_{s,3}e^{-iH_{s,3}(t+\delta t/4)\delta t/2}\nonumber\\
 	 &&\hat{S}_{s,2}e^{-iH_{s,2}(t+\delta t/4)\delta t/2}\hat{S}_{s,1}e^{-iH_{s,1}(t+\delta t/4)\delta t/2}\nonumber\\
	 &&\hat{S}_{s,0}e^{-i[H_s+H_{s,0}(t+\delta t/4)]\delta t/2}.
	\label{eq:chain-int-tro-swap}
\end{eqnarray}
Taking the first evolution step $e^{-i[H_s+H_{s,0}(t+\delta t/4)]\delta t/2}$ in Eq.~\eqref{eq:chain-int-tro-swap} as an example, we show how this individual evolution term together with the swap gate $\hat{S}_{s,0}$ is incorporated in the relevant tensors of Eq.~\eqref{eq:MPS}. The further evolution steps are applied in an essentially similar way. Noting that this evolution term has four indexes because the Hamiltonian in the exponent is actually {the} sum of two tensor products of operators in the spaces of the system and the zeroth vibrational mode, namely $H_{s}\otimes\mathds{1}_{0}$ and $A_{s}\otimes(d_0(t+\delta/4)\hat{b}_0+\mathrm{h.c.})$, the components of this individual evolution term can be written as a rank-4 tensor $M_{ij}^{i'j'}$ in which $i$ and $i'$ are the indexes of the states in the system space, while $j,j'$ are the indexes of states in the Hilbert space of the zeroth vibrational mode. The $i'$ and $j'$ indexes can be combined and regarded as the composite index for the row of the matrix of this individual evolution operator and $(i, j)$ are the columns. To apply this individual evolution operator to the MPS of Eq.~\eqref{eq:MPS}, we first contract the tensors $\Theta^{\sigma_s,\sigma_0}_{a_1} = \sum_{a_0,a_1} \Gamma_{a_0'}^{\sigma_s}S_{a_0',a_0}\Gamma_{a_{0},a_1'}^{\sigma_0}S_{a_1',a_1}$. The contracted tensor describes the effective two-site wave function of the system site and the zeroth vibrational site. Note that $\sigma_s$ and $\sigma_0$ are the indexes $i$ and $j$ that we used before. Thus,  $\Theta_{i',j'}=\Theta_{\sigma_s,\sigma_0}$. The next step is to contract the evolution term with the effective two site wavefunction $\tilde{\Theta}_{a_1}^{i'j'} = \sum_{ij}M_{ij}^{i'j'}{\Theta}_{a_1}^{ij} $. Then the two physical legs $i'$ and $j'$ are exchanged : $\tilde{\Theta}_{a_1}^{i'j'}\to\tilde{\Theta}_{a_1}^{j'i'} $. The exchange is the effect of the swap gates $\hat{S}_{s,0}$. The evolved and swapped effective wavefunction is then Schmidt decomposed: $\tilde{\Theta}_{a_1}^{j'i'}=\sum_{a_0}\tilde{\Gamma}_{a_0}^{j'}\tilde{S}_{a_0}B_{a_0,a_1}^{i'}=\sum_{a_0}\tilde{\Gamma}_{a_0}^{\sigma_0}\tilde{S}_{a_0}B_{a_0,a_1}^{\sigma_s}$ and the tensor $B_{a_0,a_1}^{\sigma_s}$ is contracted with the inverse of $S_{a_1'a_1}$ to obtain the tensor $\tilde{\Gamma}_{a_0,a_1}^{\sigma_s} = \sum_{a_1'} B_{a_0,a_1'}^{\sigma_s}(S^{-1})_{a_1',a_1}$ where $(S^{-1})_{a_1',a_1}=\delta_{a_1',a_1}S_{a_1a_1}^{-1}$. The gamma tensors $\tilde{\Gamma}_{a_0}^{j'},\tilde{\Gamma_{a_0,a_1}^{\sigma_s}}$ and singular values  $\tilde{S}_{a_0}$ then replace the previous corresponding counterparts, which completes a single evolution step.

\subsection{Scaling Analysis}
\label{sec:scaling}
The three Hamiltonian schemes introduced in Section \ref{sec:theory} have different computational costs. 
We present a qualitative scaling analysis to show the improved computational efficiency {of} the interaction-picture chain geometry (IC), compared to the chain geometry (C) scheme.
We use $d_\mathrm{IC}$, and $d_\mathrm{C}$ to denote the energy levels (local dimensions) required for the IC schemes and the C scheme. $D_\mathrm{IC}$ and $D_\mathrm{C}$ are the bond dimensions (i.e., the number of important singular values) required for the IC and C scheme. In the MPS simulations of a typical open quantum system (e.g., the spin-boson model), the most expensive step is the singular value decomposition (SVD) {for} tensors of size $(d_1D_1, d_2D_2)$, where $d_1$ and $d_2$ are the left and right local dimensions while $D_1$ and $D_2$ are the left and right bond dimensions. 
For simplicity, we assume $D_1=D_2$ and $d_1 \leqslant  d_2$. The computational cost for such a SVD step is $\mathcal{O}((d_1D)^2d_2D)=\mathcal{O}(d_1^2d_2D^3)$. In the IC scheme,  $d_1$ is the dimension of the system. For example,  $d_1=2$ for a spin. Therefore, a SVD step in the IC scheme scales as $\mathcal{O}(4d_\mathrm{IC}D_\mathrm{IC}^3)$. Such a step in the C scheme scales as $\mathcal{O}(d_\mathrm{C}^3D_\mathrm{C}^3)$ if we assume that the left and right local dimensions are identical. Section \ref{sec:res} shows that the bond dimensions in the IC and C schemes usually satisfy a linear relationship: $D_{IC} = k D_C$ where $k$ is generally larger than 1. Considering this linear relation, a SVD step in the C scheme scales as $\mathcal{O}((d_{IC}/k)^3D_{IC}^3)$. This means that, as long as $d_C^3>4k^3d_{IC}$, the IC scheme will be faster than the C scheme, at least for the SVD procedures. 
This condition is satisfied in many systems of chemical interests. An examples is the system studied in Section \ref{sec:res}, which shows that $k$ is $2\sim3$ and $d_{C}/d_{IC}\approx 8\sim10$ in the adiabatic cases.
The scaling in the star geometry (S) is higher than in the IC scheme, regardless of the interaction strength because the  entanglement (and hence the bond dimensions) grows rapidly with time (see Section \ref{sec:res}). For more general open quantum systems of interests in chemistry, the systems often involve a few electronic states (e.g., as in  electron and energy transfer systems \cite{onuchic1986some}), or can be mapped to a chain of coupled 2-state systems \cite{schroter2015exciton,holstein1959studies}, and the bath is composed of vibrational degrees of freedom with local dimensions that are much larger than the electronic degrees of freedom.

\section{Computational Details}
To demonstrate the efficiency of the method developed in section \ref{sec:theory},  we use the chain Hamiltonian Eq.~\eqref{eq:IC} in the interaction picture and the corresponding Trotterized evolution operators with swap gates (Eq.~\eqref{eq:chain-int-tro-swap}) to study a zero-bias spin-boson Hamiltonian. We compare the computational costs of this interaction-picture scheme to the chain and star geometries in the Schr\"odinger picture. The zero-bias spin-boson Hamiltonian is
\begin{eqnarray}
	H =&& \Delta\hat{\sigma}_x + \hat{\sigma}_z\otimes \int_{\Omega_0}^{\Omega_1}(\hat{a}_\omega^\dagger + \hat{a}_\omega)h(\omega)\,\mathrm{d}\omega\nonumber\\
	&& + \nonumber \int_{\Omega_0}^{\Omega_1}\hat{a}_\omega^\dagger \hat{a}_\omega \omega \,\mathrm{d}\omega.
\end{eqnarray}
where $\Delta$ is the coupling between two states $\ket{\uparrow}$ and $\ket{\downarrow}$, and $\sigma_x$ and $\sigma_z$ are the Pauli matrices. We assume that the initial state of the spin-boson model is a direct product state $\ket{\psi(t=0)} = \ket{\uparrow}_s\otimes\ket{0}_0\otimes\ket{0}_1\otimes\cdots\ket{0}_N$.  This choice follows from the initial equilibrium finite-temperature bath state in $\rho(0) = \rho_s\otimes e^{-\beta H_b}$ {($\beta=1/k_BT$)} that is mapped to the state of a zero-temperature bath by defining a temperature-dependent spectral density \cite{tamascelli2019efficient}. Following {Ref} \cite{thoss2001self}, we choose the Drude spectral density of the bath
\begin{eqnarray}
	J(\omega)=\frac{\eta \omega_{c} \omega}{\omega_{c}^{2}+\omega^{2}},
\end{eqnarray}
where $\eta$ is the coupling strength and $\omega_c$ is the characteristic frequency of the bath modes. The reorganization energy of the spin-boson model is $\frac{4}{\pi}\int_0^\infty J(\omega)/\omega\,\mathrm{d}\omega=2\eta$ because the two spin states are coupled to the bath in opposite ways through the $\sigma_z$ operator. 
The parameters used for the simulations of Section~\ref{sec:res} are  shown in Table.~\ref{tab:1} .
\begin{table}
	\centering
	\begin{tabular}{cl}\toprule
		Parameter   & Physical Quantity                                    \\\midrule
		$\Delta$      & Electronic coupling \\
		$\eta = \eta_0\Delta$\      & System-bath coupling\\
		$\omega_c = \omega_0\Delta$ & Characteristic bath frequency\\
		$k_BT = T_0\Delta$ & Bath temperature \\
		\bottomrule
	\end{tabular}
	\caption{The parameters. $\eta_0$, $\omega_0$ and $T_0$ are unitless quantities that scale the system-bath coupling, the frequency of the bath and the temperature of the bath, respectively.}
	\label{tab:1}
\end{table}
The ordering of sites in the star geometry plays a crucial role in the growth of the bond dimensions with time. In the star geometry, we order the bath sites based on the absolute values of their frequencies, with the lowest-frequency mode closest to the system (the spin). This ordering is used widely in MPS simulations of vibronic systems \cite{ren2018time,borrelli2021finite,xie2019time}.

\section{Results and Discussions}
\label{sec:res}
\subsection{Spin-Boson Model in Strong-coupling Regimes}
We now show the efficiency of the interaction-picture chain-geometry scheme (IC) using the spin-boson Hamiltonian in a strong coupling regime at moderate to high temperatures. The dimensionless characteristic frequency $\omega_0$ {is} set to be $0.25,1.0$ or $4.0$, corresponding to the adiabatic, intermediate and non-adiabatic regimes, respectively. The population dynamics at low temperatures, or in the weak coupling regimes,  are not studied here because they are relatively easy to simulate due to the low excitation of bath modes and the low entanglement.

\subsubsection{Adiabatic Regime: $\omega_c\ll \Delta$}
In the adiabatic regime, the electronic coupling $\Delta$ is large compared to the characteristic frequency $\omega_c$ of the bath. We use a small characteristic frequency $\omega_0=0.25$ and a moderate temperature $T_0= 1.0$ so that $\omega_c=\omega_0\Delta=0.25\Delta$ and $T = T_0\Delta=\Delta$. We plot the computed population evolution of the spin $\ket{\uparrow}$ state in Fig.~\ref{fig:chainAndStar}. The dynamics are calculated by the three different schemes: the interaction-picture chain geometry (IC), the Schr\"odinger-picture chain geometry (C), and the star geometry (S). Each scheme  uses various cutoffs for the dimension of the Hilbert space (local dimension) of the bath modes to show the convergence trends. Here and below, the threshold for the singular values in the numerical simulations is $10^{-3}$. That is, the singular values smaller than $10^{-3}$ are discarded. The maximum number of singular values is restricted to 1000. We also perform the simulations with a tighter threshold $10^{-4}$ (not shown).
No significant difference between the simulation results with the two different thresholds was found, which indicates the results are converged with respect to the threshold of $10^{-3}$ for singular values. The dimensionless time step $\delta t$ in all of the simulations is $5\times 10^{-2}$, unless otherwise noted.

\begin{figure}
	\centering\includegraphics[width=\linewidth]{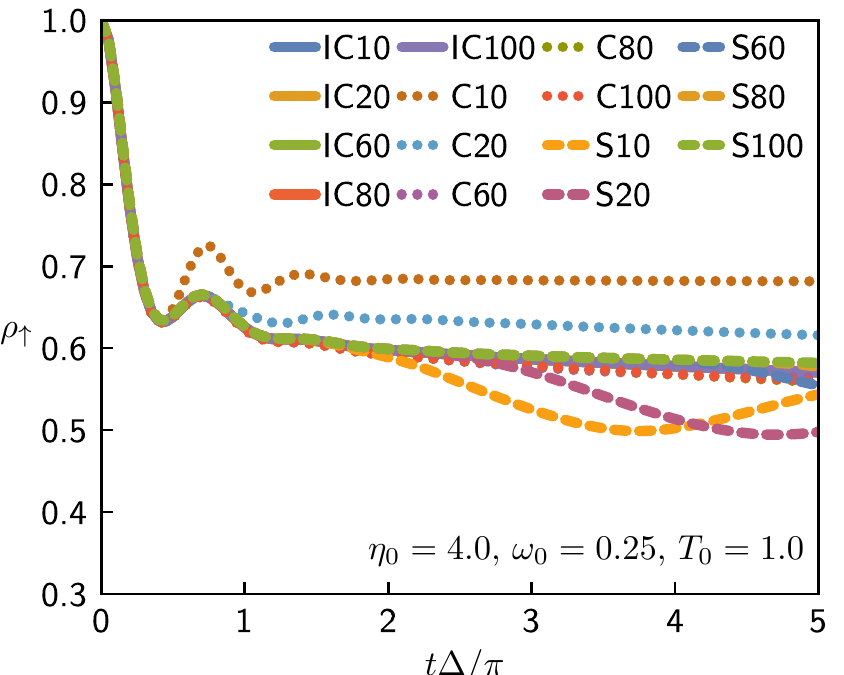}
	\caption{Evolution of the $\ket{\uparrow}$ state population. ``C", ``IC", and ``S" denote the chain geometry in the Schr\"odinger picture, the chain geometry in the interaction picture, and the star geometry in the Schr\"odinger picture, respectively. The numbers 10, 20, etc. are the maximum quantum numbers (cutoff) of bath modes.}
	\label{fig:1}
\end{figure}

As shown in Fig.~\ref{fig:1}, the simulations for the chain geometry or the star geometry (denoted C and S)  in the Schr\"odinger picture are different from the converged results obtained in the interaction-picture chain geometry (denoted IC),  unless a sufficiently large dimension for the bath-mode Hilbert space (the local dimension) is used.
We find that a cutoff of 10 for the local dimension is sufficient for the IC scheme to converge and a cutoff of 60-80 is required to ensure convergence in the C scheme. The cutoff of 80 needed for the S scheme is much less favorable than both IC and C schemes. 

\begin{figure}
	\centering\includegraphics[width=.8\linewidth]{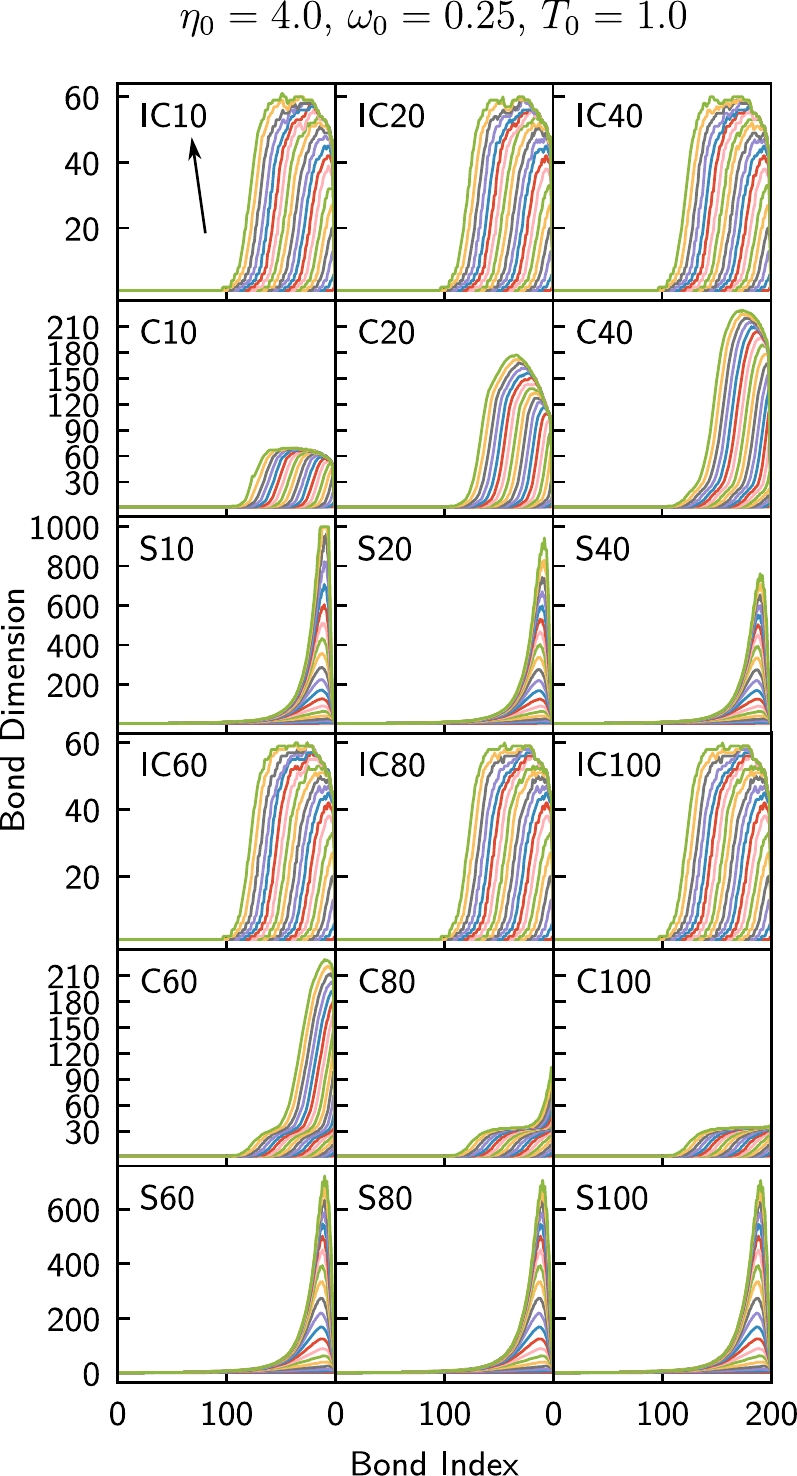}
	\caption{The growth of bond dimensions in the IC, C, and S schemes. The $x$-axis is the index of the bond between every neighboring  two sites on the MPS. The last bond (index 200) is the bond between the spin and the first bath mode. In each panel, curves in different colors represent the bond dimensions at different times and the bond dimensions grow from zero at the time zero to a larger value at the time $t\Delta/\pi=5$. The arrow points to the direction of the growth (also the direction of time).}
	\label{fig:2}
\end{figure}

\begin{figure}
	\centering\includegraphics[width=\linewidth]{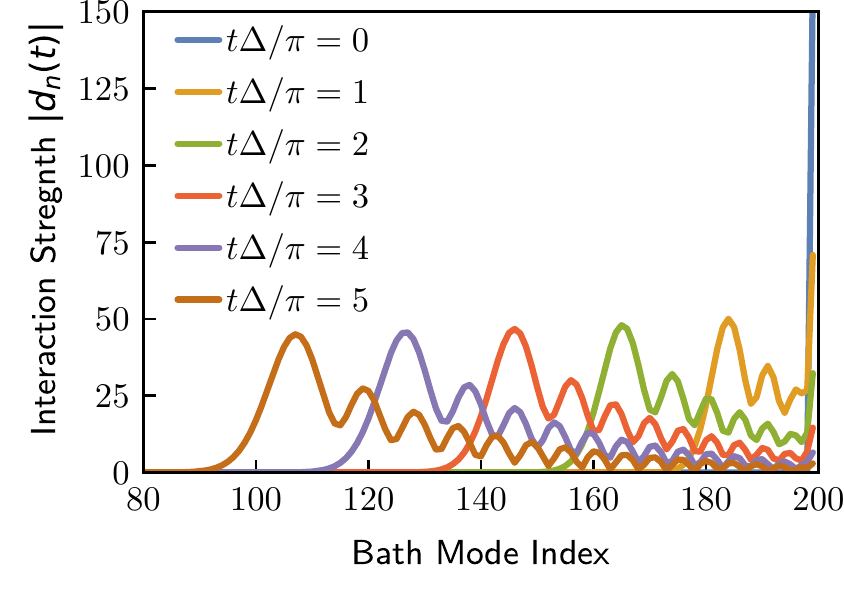}
	\caption{The absolute value of the complex-valued interaction strengths $d_n(t)$ where $n$ is the index for the bath modes. The blue line is $d_n(t=0)$ and the spin only interacts with the first mode (index 200) at this moment. The rest of the colors are the interaction strength at later times. The time-dependent $|d_n(t)|$ exhibits a traveling-wave pattern and at different moments the largest interaction strength occurs at different modes. Qualitatively speaking, the large interaction strength at $t=0$ is distributed into multiple modes at later times. This explains why every mode in the interaction-picture is less excited and why the interaction picture scheme can still enjoy the slow growth of  bond dimensions.}
	\label{fig:dk}
\end{figure}

To further explore the differences among the three schemes, Fig.~\ref{fig:2} shows the growth of bond dimensions during the simulations of Fig.~\ref{fig:1}, with different cutoffs for the local dimensions.
The results in the adiabatic regime show that the growth in the bond dimensions has different patterns in the three schemes. The slowest bond-dimension growth occurs in the interaction-picture chain geometry (IC) and the Schrodinger-picture chain geometry (C). The fastest growth occurs in the star geometry. Compared to the Schr\"odinger picture chain geometry, the interaction-picture chain scheme exhibits a similar growth pattern of the bond dimensions, but involves much smaller matrices as discussed in Section~\ref{sec:intro}. This slow growth of bond dimensions occurs because the time-dependent interactions in the IC scheme exhibit a traveling-wave pattern as shown in Fig.~\ref{fig:dk}. Furthermore, the interactions are delocalized among multiple modes, and each interaction is weaker than the system-bath interaction in the Schr\"odinger-picture chain.

\subsubsection{Intermediate Regime: $\omega_c\sim\Delta$}
\begin{figure}
	\centering\includegraphics[width=\linewidth]{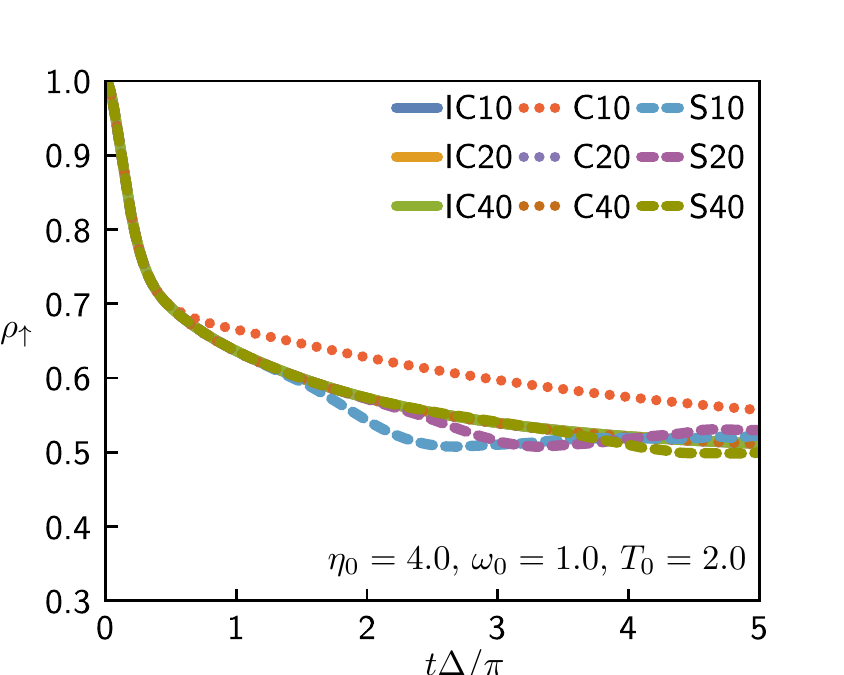}
	\caption{The evolution of state $\ket{\uparrow}$ population. The threshold for singular values is $10^{-4}$. The dimensionless time step $\delta t$ is $5\times10^{-3}$.}
	\label{fig:3}
\end{figure}

In the intermediate regime,  the magnitude of the characteristic frequency $\omega_c$ of the bath modes is comparable to the electronic coupling. We set $\omega_0=1.0$ so $\omega_c=\omega_0\Delta=\Delta$. If the same temperature value $T_0=1.0$ is chosen as in the adiabatic case, we expect that the occupations in the bath modes will be lower than in the adiabatic case because a higher frequency ($\omega_0=1.0$) is set in the intermediate regime. To make the comparisons between the different schemes more distinct, the dimensionless temperature is set to $T_0=2.0$, which is  higher than the temperature studied in the adiabatic regime. This higher temperature is more costly for the simulations than the low-temperature regimes because high-lying states of the bath are populated. The computed population dynamics are shown in Fig.~\ref{fig:3}. 
In the IC scheme, a converged result is found with a small cutoff for the local dimensions of the bath modes. This result is similar to the situation in the adiabatic regime. However, the C and S schemes in the intermediate coupling regime converged more rapidly with respect to local dimensions than in the adiabatic regime.  This milder requirement for local dimensions occurs because the frequencies of the modes are comparable to the off-diagonal coupling of the spin (the system) and fewer energy levels of the modes are excited than in the adiabatic (small frequencies) case. Note that the temperature is doubled compared to the adiabatic case of Fig.~\ref{fig:1}, while the frequency is quadrupled. 
Although the requirement for local dimensions is weaker for all the schemes compared to the adiabatic case, the IC scheme considerably outperforms the C and S schemes, benefiting from the reduced matrix and tensor sizes by requiring a smaller cutoff of local bath dimensions in the intermediate regime. The IC scheme can become even more efficient as the temperature grows
.  A considerable number of electron-transfer \cite{tamura2012quantum}, singlet-fission \cite{fujihashi2016fluctuations}, and excitation-energy-transfer (EET) systems \cite{ishizaki2009unified,huo2011theoretical} fall in this intermediate regime, where the magnitude of the electronic coupling is comparable to vibrational frequencies. We expect that the IC scheme can accelerate MPS simulations for chemical systems of interest, especially light-harvesting systems where relatively low-frequency bath modes are important \cite{christensson2012origin}.

\begin{figure}
	\centering\includegraphics[width=\linewidth]{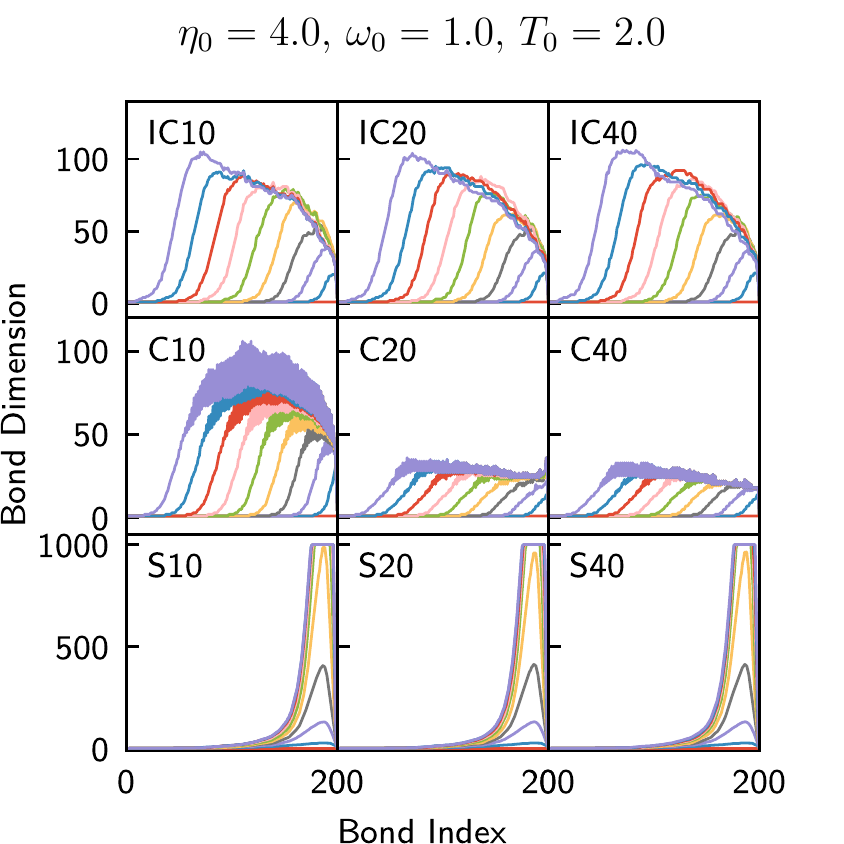}
	\caption{The growth of bond dimensions in the Schr\"odinger-picture chain geometry, the interaction-picture chain geometry, and the Schr\"odinger-picture star geometry, in the intermediate regimes. 
	}
	\label{fig:4}
\end{figure}

The entanglement of each bond is explored in Fig.~\ref{fig:4}, which shows the growth of bond dimension for the simulations of Fig.~\ref{fig:3}.  
As in the adiabatic cases, the IC and C schemes have similar growth patterns of bond dimensions as long as the results in the C scheme have converged (e.g., C20 and C40), although the IC scheme has slightly faster growth of bond dimensions (twice as fast as in C).  The similarity of the C and IC scheme on the growth pattern of bond dimensions shows the similar kinds of energy flow through the bath chain, which is due to the traveling-wave pattern of the time-dependent interactions in the IC scheme (see Fig.~\ref{fig:dk}).   For the S scheme, although the dynamics are converged, the bond dimensions still grow rapidly because of the intrinsic structure of the MPS in the star geometry. This finding indicates that the structure of MPS in the S scheme is not optimal, since the spin interacts with all of the modes at any time.

\subsubsection{Non-Adiabatic Regime: $\omega_c\gg\Delta$}
\begin{figure}[H]
	\centering\includegraphics[width=\linewidth]{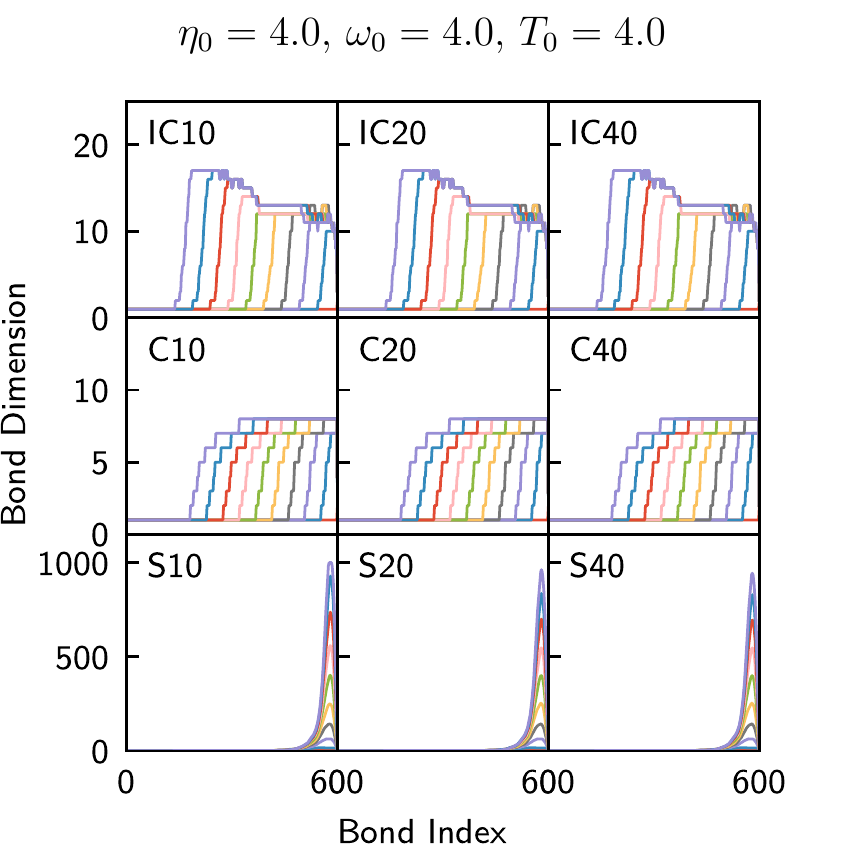}
	\caption{The growth of bond dimensions  in the intermediate regimes for the Schr\"odinger-picture chain geometry, the interaction-picture chain geometry, and the Schr\"odinger-picture star geometry.}
	\label{fig:6}
\end{figure}

In the non-adiabatic regime, we set the characteristic frequency $\omega_c$ of the bath modes to be $\omega_c=4\Delta$, and the temperature to be $T=T_0\Delta=4.0\Delta$. In this regime, the simulations do not require too many bath-mode energy levels for the calculations to converge. No significant deviations are shown in the IC and C schemes, regardless of the local dimension cutoffs, except for the S scheme (which shows strong deviations due to the rapid growth of entanglement for the star geometry, see Fig.~\ref{fig:6}).
Although the IC and C schemes do not show significant differences as we vary local dimensions (for the temperature used), we expect IC scheme can benefit in speed from the smaller matrices and the smaller local dimensions in the IC scheme. The smaller matrices in the IC scheme may also improve the speed of numerical simulations for low-temperature regimes where the classical Marcus rate formula does not apply and the non-adiabatic rate is sensitive to details of the spectral density \cite{egger1994quantum}.

\begin{figure}[H]
	\centering\includegraphics[width=\linewidth]{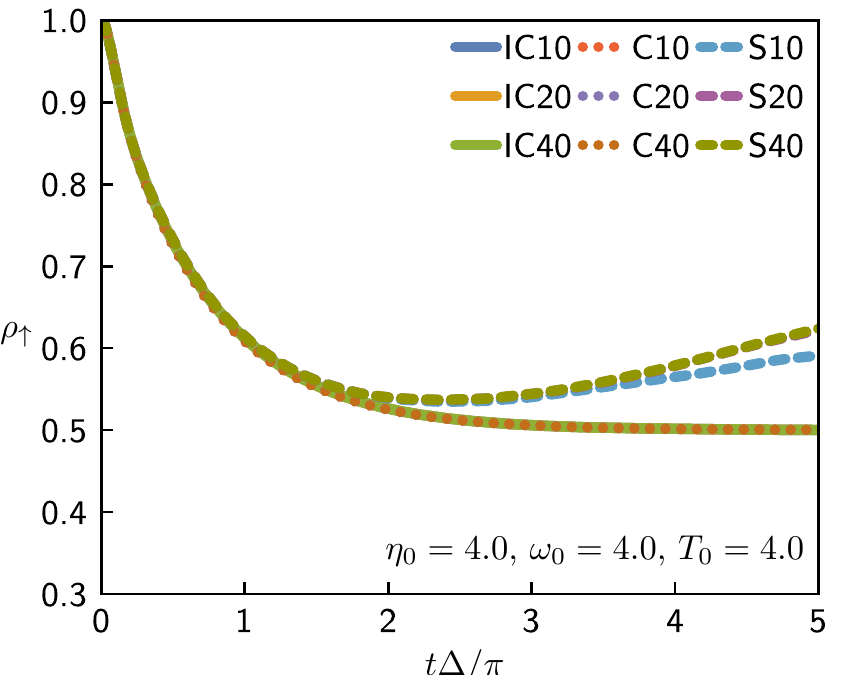}
	\caption{The time evolution of the state $\ket{\uparrow}$ population on  in the non-adiabatic regimes. The threshold for singular values is $10^{-3}$. The dimensionless time step $\delta t$ is $1.25\times10^{-2}$. The dynamics in the IC and C schemes are converged using a small local dimension (10), while the S scheme did not yield converged results regardless of the number of vibrational levels. This is because the bond dimension in the S scheme grows extremely rapidly, and the current threshold ($10^{-3}$) for the singular values is not enough for accuracy.}
	\label{fig:5}
\end{figure}

\subsection{Computational Efficiencies in the Interaction Picture}
The numerical simulations described here show that the IC scheme requires fewer bath energy levels than the C scheme to obtain converged results in the adiabatic, strong-coupling, high-temperature regimes.
Using the scaling relations of Section \ref{sec:scaling}, we analyze the simulations of Fig.~\ref{fig:2} (i.e., the adiabatic spin-boson model). We find that $D_\mathrm{IC}\approx 2D_\mathrm{C}$, $d_\mathrm{C} = 80$, $d_\mathrm{IC}=10$. As a result, the scaling of a single SVD step in the C scheme is $\mathcal{O}(80^3D_\mathrm{C}^3)=\mathcal{O}(51200D_\mathrm{C}^3)$ and the scaling in the IC scheme is $\mathcal{O}(4\cdot 10 \cdot 8D_\mathrm{C}^3)=\mathcal{O}(320D_\mathrm{C}^3)$. The findings indicate that the IC scheme is roughly 1600 times faster than the C scheme for the SVD operations. If we assume $D_\mathrm{C} \approx 2D_\mathrm{IC}$,  the C scheme is faster than the IC scheme ($\mathcal{O}(4d_\mathrm{IC}D_\mathrm{IC}^3)<\mathcal{O}(d_\mathrm{C}^3D_\mathrm{C}^3)$) only if $d_\mathrm{C}^3 < 32d_\mathrm{IC}$. This relation means that if the local dimension for a bath mode is larger than $6$ ($\approx \sqrt{32}$), one should use the interaction-picture chain geometry (IC scheme). 

For the star geometry (the S scheme), the SVD step also scales as $\mathcal{O}(4d_{S}D_{S}^3)$, but $D_S$ is often a large number. As a result, the S scheme is always slower than the IC scheme.

\section{Conclusions and Outlook}
\label{sec:conclusion}
By transforming the chain Hamiltonian to the interaction picture, we {developed} a numerical approach that outperforms the conventional chain-geometry and star-geometry approaches in the moderate-to-strong coupling regimes at intermediate-to-high temperatures. The efficiency of this approach is demonstrated by simulating the spin-boson model and comparing the growth of bond dimensions to the conventional approaches. The results indicate that working in the interaction picture gains increased efficiency in simulating open quantum systems with boson environments using the t-DMRG method, especially when the population of the excitated boson states are high. Possible extensions of this interaction-picture approach could include anharmonic baths, the combination with the optimized-boson-basis (OBB) or the local-basis-optimization (LBO) approach \cite{guo2012critical,schroder2016simulating}  \cite{brockt2015matrix} to further accelerate singular value decomposition and the use of the Polaron transformation \cite{silbey1984variational,zueco2019ultrastrongly,shi2018ultrastrong,he2018improved} to reduce entanglement in the interaction picture.

\begin{acknowledgments}
	This material is based upon work supported by the National Science Foundation under Grant No. 1925690.
\end{acknowledgments}

\bibliography{references}

\end{document}